\documentclass[aps,showpacs,floatfix]{revtex4}
\usepackage{graphicx}
\newcommand{\be}{\begin{eqnarray}}
\newcommand{\ee}{\end{eqnarray}}
\def\beq{\begin{equation}}
\def\eeq{\end{equation}}

\begin{document}
\title{Power spectrum for critical statistics:\\
A novel spectral characterization of the Anderson transition}
\author{Antonio M. Garc\'{\i}a-Garc\'{\i}a}
\affiliation{Physics Department, Princeton University, Princeton, New Jersey 08544, USA}
\affiliation{The Abdus Salam International Centre for Theoretical
Physics, P.O.B. 586, 34100 Trieste, Italy}
\begin{abstract}
We examine the power spectrum of the energy level fluctuations  
 of a family of critical power-law random banded 
 matrices with properties similar to those of a disordered
 conductor at the Anderson transition. 
It is shown both analytically and
 numerically  that the Anderson transition  
is characterized by a power spectrum which presents
 $1/f^2$ noise for small frequencies but $1/f$ noise for larger frequencies.
For weak diagonal disorder the analysis 
 of the transition region 
between these two power-law limits provides with an accurate estimation of the  
 Thouless energy of the system. 
As disorder increases the Thouless energy looses its meaning and the 
 power spectrum presents a $1/f^2$ decay  up to 
frequencies related to the Heisenberg time of the system.
Finally we discuss under what circumstances these findings 
 may be relevant in the context of non-random Hamiltonians.

\end{abstract}
\pacs{72.15.Rn, 71.30.+h, 05.45.Df, 05.40.-a} 
\maketitle
\section{Introduction}
Level statistics is a powerful tool to investigate the properties of quantum complex systems. 
The spectrum is basis independent and
typically easier to access either numerically or experimentally than the eigenfunctions.
The analysis of the level statistics is usually carried out in two steps. First the  
 spectrum is properly unfolded, namely, by extracting the mean level density, the original spectrum is transformed 
 into one with a mean level density equal to the unity. In a second stage the unfolded spectrum is analyzed
 by evaluating different spectral correlators.  
Two popular choices are the level spacing distribution $P(s)$ (the probability of
 having two eigenvalues at a distance $s$) for short range correlations and the number variance 
$\Sigma^{2}(L)=\langle L^2 \rangle - {\langle L \rangle}^2$ (which measures the deviations of 
 the number of eigenvalues in an interval $L$ from its mean value) 
for long range correlations. In certain situations these spectral correlators present striking universal
features. For instance,
 in the context of deterministic Hamiltonians the celebrated Bohigas-Giannoni-Schmit conjecture \cite{oriol} states 
the level statistics of a deterministic quantum system whose classical 
counterpart is fully chaotic depends not on the microscopic details of the Hamiltonians but
only on the global symmetries of the system 
and are identical to those of a random matrix with the same symmetry,
 usually referred to as Wigner-Dyson statistics (WD) \cite{mehta}. 

Remarkably the same WD statistics also describes \cite{efetov} the spectral correlations of a disordered system in the metallic limit.  
 In the strong disorder limit localization sets in, the spectrum is not correlated
 and the level statistics is universally described by Poisson statistics. For deterministic
 systems the same statistics is generic of systems whose classical dynamics is 
 integrable \cite{tabor}. From a practical point of view, universality is usually tested by comparing correlators like
$P(s)$ or $\Sigma^2(L)$ of a specific system with the universal predictions of WD or Poisson 
statistics.

Recently a different spectral characterization based on techniques of 
time series was introduced in the context of quantum chaos \cite{rel,rel1}.      
In \cite{rel,rel1} the unfolded energy spectrum is formally considered as a discrete signal and the energy levels 
 are interpreted as a time series. Specifically they investigate the 
 power spectrum,
\be
 S(k)=\left| {1\over \sqrt{N}} \sum_{n=1}^N \delta_n \exp{\left(
{-2\pi i k n\over N} \right)} \right|^2 \; , 
\ee 
of the sequence $\delta_n$
\be 
\delta_n =
\sum_{i=1}^n\left(s_i-{\bar s}\right)=\epsilon_{n+1}-\epsilon_1-n,
\ee  
 where $s_i=\epsilon_{i+1}-\epsilon_i$, $\epsilon_i$ is ith unfolded 
 eigenvalue and $N$ is the length of the series. The correlator $\delta_n$ thus gives 
the deviation of the ith nearest neighbor spacing $s_i$ from its mean value ${\bar s}$
which by definition is the unity for unfolded eigenvalues. 
 
It was found that the $S(k)$ for systems whose classical 
 counterpart is completely chaotic or integrable  
has a universal power-law form with an exponent depending on the classical dynamics:
 $S(k) \sim 1/k$ for chaotic and $S(k) \sim 1/k^2$ for integrable 
 motion.

Universality in the spectral correlations has also a counterpart in the eigenfunctions properties. Thus Poisson 
 statistics is associated with exponential localization of the eigenfunctions and WD statistics is typical 
 of systems in which the eigenstates are delocalized through the sample and can be effectively represented by a superposition  of plane waves with random phases. 

Despite its robustness, these universal features are restricted to long 
 time scales (related to energy scales of the order of the mean level spacing) such that
an initially localized wave-packet has already explored the whole phase space available. In other 
 words, universality is related to certain ergodic limit of the quantum dynamics. For shorter time scales 
 the system has not yet relaxed to the ergodic limit and deviations from universality are expected.
For finite disordered systems this scale is given by
the dimensionless conductance $g = E_c/\Delta$ ($E_c$, the Thouless energy, is a scale of energy associated 
with the classical diffusion time through sample and $\Delta$ is the mean level spacing) which roughly 
 speaking gives the number
 of eigenvalues whose  
spectral correlations are universally described by WD statistics. 

Striking universal features not related to any ergodic limit (they persist beyond the 
 mean level spacing scale) have
 also been observed in a disordered system at the metal-insulator transition
also referred to as Anderson transition (AT). 
It is by now well established that a disordered system with short range
hopping in more than two dimensions undergoes an AT \cite{anderson,wegner} at the center of the band for a
critical amount of disorder (for critical  
 we mean a disorder such that if increased all the 
 states in the band become exponentially localized). 
Since 
 the dimensionless conductance $g \sim 1$ is about the unity
 at the 3D (or 4D) AT, the level statistics for eigenvalues 
 separations larger than the mean level spacing describes truly dynamical features of the system.
By universality we mean that the level statistics do  not depend on boundary conditions,
 shape of the system or the microscopic details of the disordered potential though some features may depend
 on the dimensionality of the space.  

Signatures of the AT are found in both the level statistics and the 
 eigenfunctions.
Systems belonging to this new universality  class have 
multifractal eigenstates.
Intuitively multifractality 
 means that the eigenstates have structures at all scales.
In a more formal way multifractality is defined through the anomalous scaling of the eigenfunctions 
moments ${\cal P}_q=\int d^dr |\psi({\bf r})|^{2q}$ with respect to the sample size $L$ as ${\cal P}_q\propto 
L^{-D_q(q-1)}$, where $D_q$ is a set of different exponents describing the AT \cite{aoki}. 
 
Level statistics
at the AT (commonly referred to as 
'critical statistics \cite{kravtsov97}) is intermediate between 
 WD and Poisson statistics. Although a formal definition is 
 still missing, typical features of critical statistics include:
 scale invariant spectrum \cite{sko},
 level repulsion and  linear number 
 variance ($\Sigma_2(L) \sim \chi L$) \cite{chi} as for a insulator ($\chi = 1$) but
 with a slope $\chi < 1$ ($0.27$ for the 3D Anderson transition). 
Similar spectral properties 
has also been found 
 in random matrix models
 based on soft confining potentials 
 \cite{log}, effective eigenvalue distributions \cite{Moshe,ant4} related to the Calogero-Sutherland model \cite{calo} 
at finite temperature
 and random banded matrices with power-law decay \cite{evers}. 
 The latter is specially interesting since an AT (for the case $1/r$ decay) has been analytically established
by mapping the problem onto a non linear $\sigma$ model.
 

 In this paper we propose an alternative spectral characterization of the Anderson transition  
 based on the analysis of the power spectrum 
$S(k)$ introduced above. We shall also see this technique provides with an accurate
 way to locate the Thouless energy of a disordered system. Finally we will discuss the 
relevance of our findings in the context of non random system. It will shown that $S(k) \sim 1/k^2$ 
is not always related to integrable classical motion. Consequently a precise classificatory scheme 
 based on $S(k)$  must have into account other features of $S(k)$ besides the exponent of the power-law decay.

The organization of the paper is as follows. In the next section 
 the model to be investigated is introduced. The 
 power spectrum $S(k)$ is evaluated both analytically and numerically 
for a broad range of parameters in section three and four. From these results we 
present a novel spectral characterization of
 the Anderson transition based on the  analysis of $S(k)$. We will also argue that this technique
 can be utilized to detect the Thouless energy in a disordered system. 
 Finally in section five we discuss in what situations our findings may be relevant 
for non-random quantum systems.

\section{The model}
In this section we evaluated both analytically and 
 numerically the power spectrum $S(k)$ of a critical random banded 
 matrix associated to 
 critical statistics.

Unlike WD or Poisson statistics, 
 critical statistics is not parameter free. 
Together with universal
features such as scale invariance, level repulsion and
  linear number variance $\Sigma^2(L) = \chi L~~L \gg 1$,  there are also 
 system dependent features as the numerical value of the slope of the number 
variance $\chi$. For instance, for short 
 range Anderson models, this slope depends on the Euclidean
 dimension $d$ of the sample.
Thus for the lower critical dimension $d=2+\epsilon$ ($\epsilon > 0$) \cite{wegner},
  $\chi \sim \epsilon \ll 1$. In the 
 opposite limit $d \gg 1$, $\chi \leq 1$ is close to the unity similar to the case of an insulator.

In this letter, instead of studying directly these short range Anderson model, we will focus on 
certain generalized random matrix models 
which has been shown to reproduce 
critical statistics \cite{kravtsov97} with great accuracy.  An advantage of these models is 
 that exact 
analytical solutions are available in a certain region of parameters \cite{Moshe,evers,log}. 
\begin{figure}[t]
\centering
\includegraphics[width=0.9\columnwidth,clip]{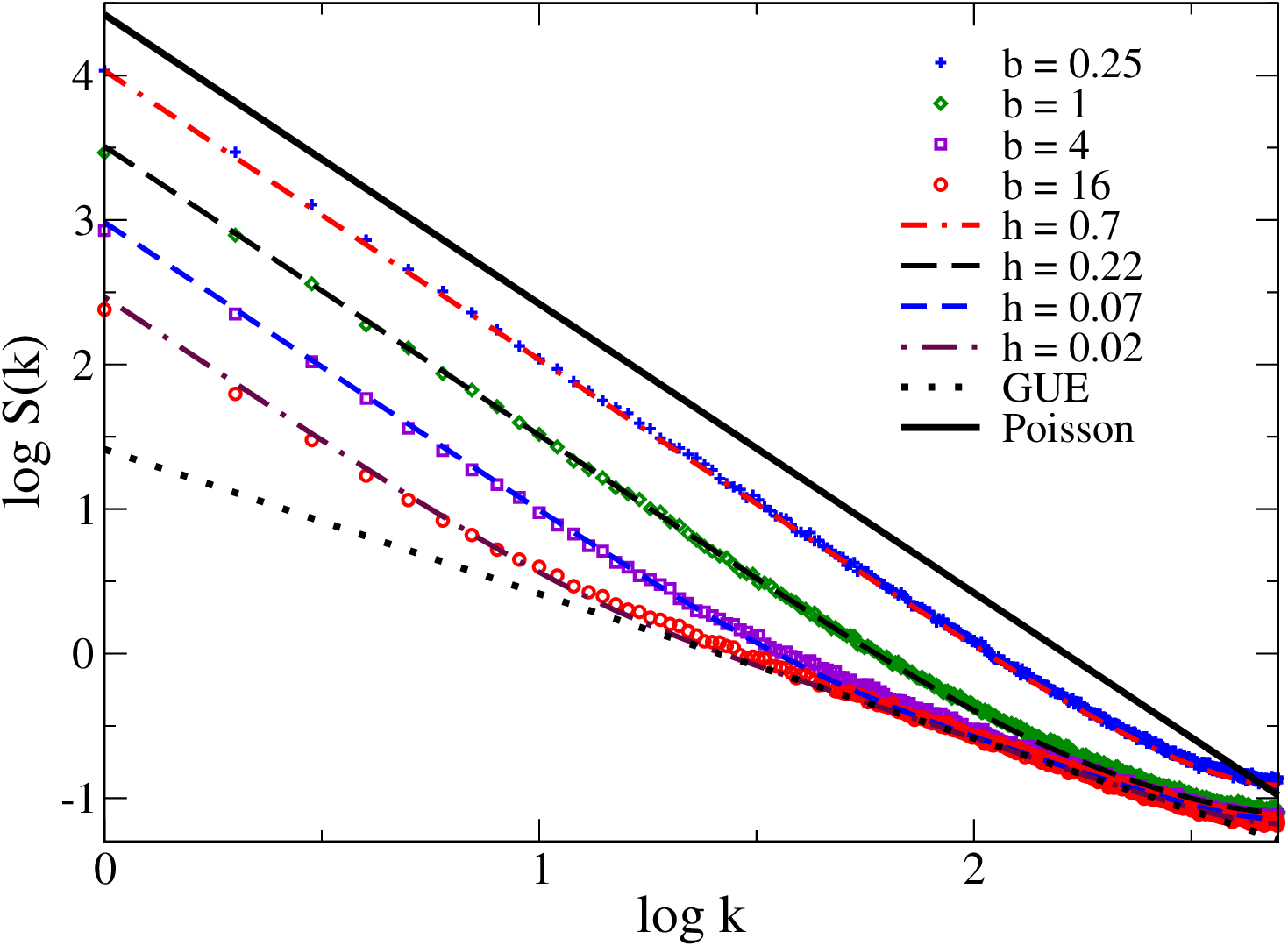}
\caption{Power spectrum $S(k)$ 
as a function of $k$. Symbols represent the numerical results (the matrix size is $3000$ and the 
 number of eigenvalues considered is $N=1024$ around the center of the band) 
for the critical random 
 banded model Eq.\ref{e4} for different bandwidth $b$, the power spectrum was evaluated
from Eq. \ref{sk}. 
Lines represent the analytical prediction of critical
statistics Eq. \ref{sfi} with the TLCF given by Eq. \ref{r23}. For all $b$ the agreement between 
 theory and numerics is excellent.
As predicted for $b \gg 1$ we observe two different regions: $N/k \gg 2\pi b$
 corresponding with $S(k) \sim 1/k^2$ and $ N/k \ll 2 \pi b$ with $S(k) \sim 1/k$ similar to 
 the prediction of WD (GUE). However for $b \ll 1$, $S(k) \sim 1/k^2$ for almost all $k$.}
\label{f1}
\end{figure}

We investigate the ensemble of
random complex Hermitian matrices $\hat H$. 
The matrix elements $H_{ij}$ are independently distributed
Gaussian variables with zero
mean $\langle H_{ij}\rangle=0$ and variance 
\begin{equation}
\label{e4}
\langle |H_{ij}|^2\rangle =\\
\left[1+{1\over b^2}{\sin^2(\pi (i-j)/N)\over(\pi/N)^2}\right]^{-1}\ ,
\end{equation}
For any value of the bandwidth
$0<b<\infty$, the spectral correlations are given by 
 critical statistics and the eigenvectors are multifractal exactly as 
 at the conventional Anderson transition in $2<d<\infty$ \cite{evers}. The limit $b \rightarrow \infty$ corresponds with 
the standard Gaussian Unitary Ensemble (GUE) of random matrices.  
The region $b\gg 1$ (weak diagonal disorder, $\chi \ll 1$) corresponds
with  $d=2+\epsilon$ ($\epsilon\ll 1$) and the $b\ll 1$ limit 
with $d\gg 1$ and $\chi \leq 1$ (strong diagonal disorder).
For Hermitian matrices these two limits are accessible to analytical 
 techniques\cite{evers,kra2005}. Here we do not discuss the details
 of these calculations but just enumerate certain results we will 
 use later on in the calculation of the power spectrum $S(k)$. 
 
For $b \gg 1$ the level statistics 
can be rigorously investigated after mapping the random banded matrix onto a   
 supersymmetry sigma model. It can be shown \cite{evers} that in this limit
the connected part of two level correlation function (TLCF) is given by,
\be
\label{r21}
R_2(s)=\frac{\langle\rho(s/2)\rho(-s/2)\rangle}{\langle \rho(0) \rangle^2}
- 1 = \delta(s) - \frac{1}{16 b^2}\frac{\sin^2(\pi s)}{\sinh^2 (\pi s/4b)}
\ee 
where $\rho(s=E/\Delta)=\sum_i \delta (s_i -s)$ 
is the spectral density in units of the mean level spacing $\Delta = 1/\langle \rho(0) \rangle$
and brackets stand for ensemble average.
 
For $b \ll 1$,  various spectral correlators can also by calculated explicitly 
by a recently developed virial expansion around the Poisson limit 
\cite{kra2005} (for a more heuristic approach see \cite{evers}).

The intermediate 
 region of $b \sim 1$ is not yet accessible to analytical techniques.  
However there is a closely related random matrix model \cite{Moshe} which is exactly solvable 
 for any $b$ and which has the same TLCF (to leading order in $b$) in the two regions ($b \ll 1, b\gg 1$)
 above discussed.
 Its joint probability
distribution is given by
\be
\label{kl1}
P(H,b)=\int dU e^{-\frac{1}{2}{\rm Tr} HH^{\dag}} e^{-\frac{b}{2}
{\rm Tr}{[U,H][U,H]^{\dag}}}.
\ee
Here, the $N\times N$ matrices  $H$ and $U$ are Hermitian and
Unitary, respectively, and
the integration measure $dU$ is the Haar measure. 
Despite its complicated form, it can be shown \cite{Moshe,ant4} that the joint distribution
 of eigenvalues of $H$ is equal to the diagonal element of the density matrix of a system of free spinless 
 fermions at finite temperature $T=1/b$ confined in an harmonic potential.
By using elementary statistical mechanics techniques it can be shown that 
the TLCF for arbitrary $b$ is given by,
\be
\label{r23}
{R_2}(s) = \delta(s)- \left 
(
\int_{0}^{\infty}\frac{\cos(\pi st/\rho(0))}
{\rho(0)}\frac{1}{{1+z e^{t^2}}}dt \right)^2
\ee      
where $h = 1/2\pi b$,$z = \frac{1}{e^{1/h}-1}$ and $\rho(0) = \int_{0}^{\infty}\frac{1}{1+ze^{t^2}}$.
We shall use this expression in the analytical evaluation of the power spectrum  for 
 intermediate values of $b$ and then check its validity by carrying out
  numerical simulations of the random banded model Eq. \ref{e4}.

\vspace {1cm}
\begin{figure}[ht]
  \hfill
  \begin{minipage}[t]{.45\textwidth}
      \includegraphics[width=\columnwidth]{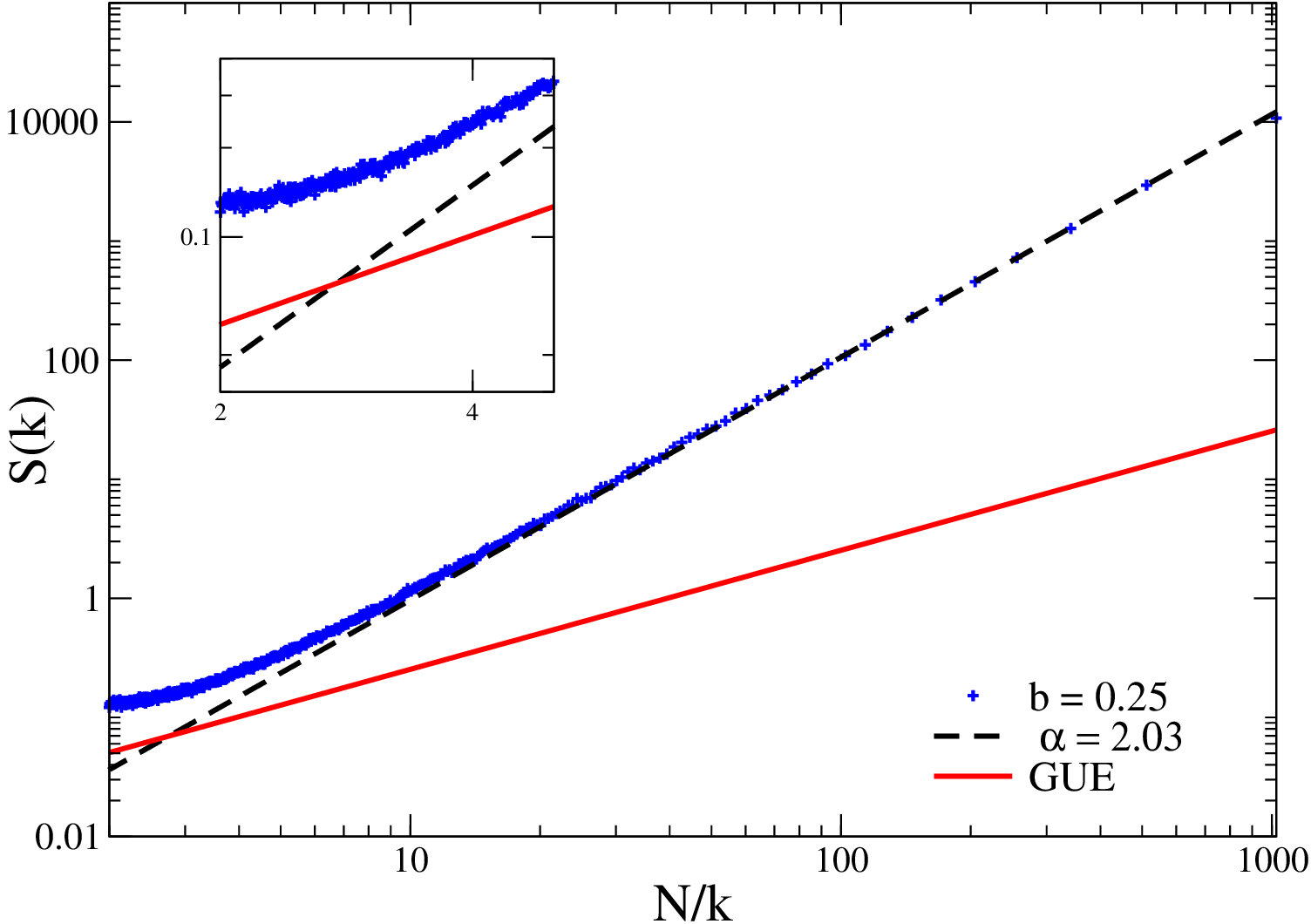}
      \label{nf0ps1}
  \end{minipage}
  \begin{minipage}[t]{.45\textwidth}
      \includegraphics[width=\columnwidth]{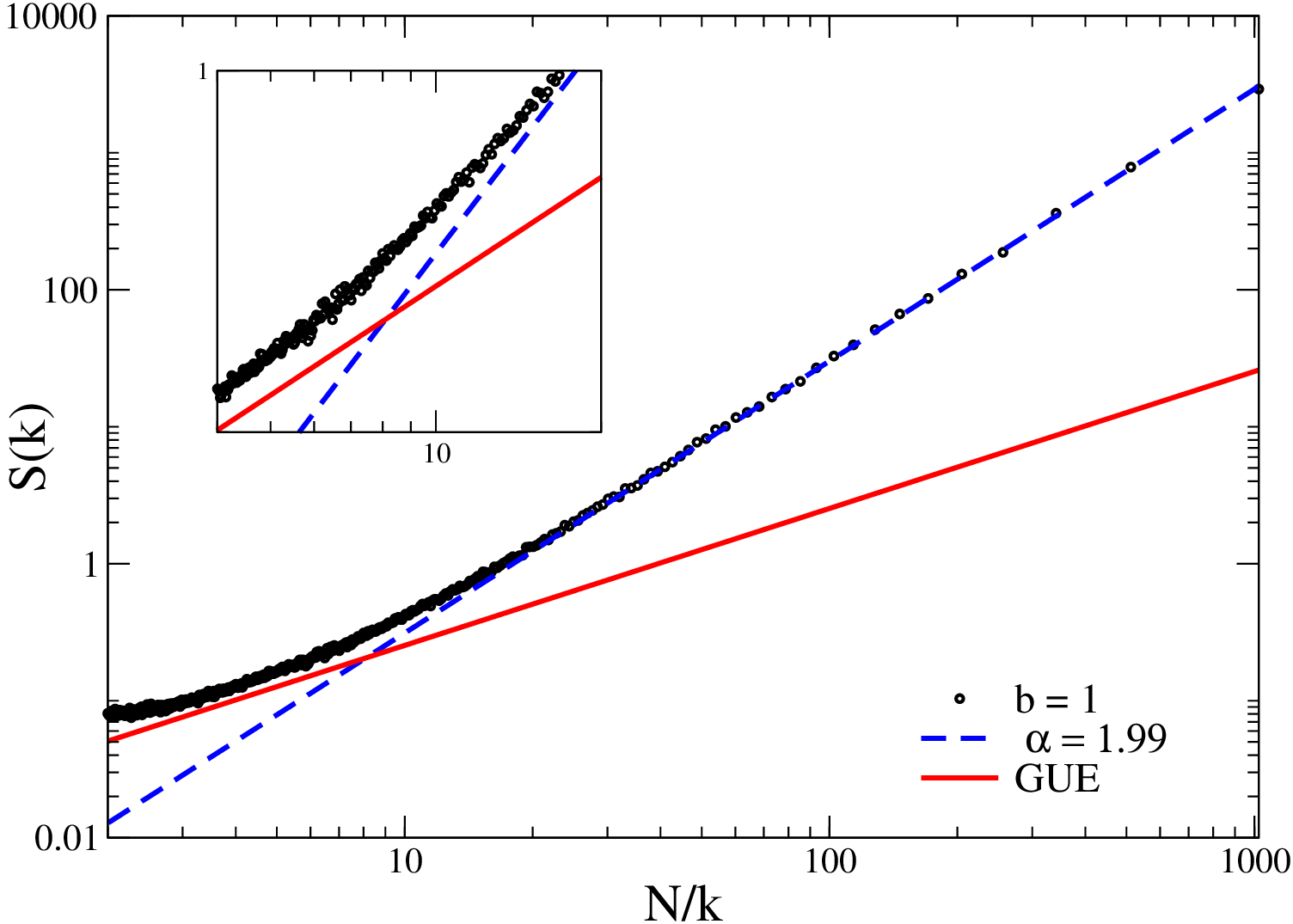}
      \label{f2}
  \end{minipage}
     \caption{Power spectrum $S(k)$ versus $N/k$ in the limit $b \leq 1$.
 Crosses and circles represent the numerical results (the matrix size is $3000$ and the 
 number of eigenvalues considered is $N=1024$ around the center of the band) 
for the critical random 
 banded model Eq. \ref{e4} for $b = 0.25,1$ respectively. The  power spectrum was evaluated
from Eq. \ref{sk}. The dashed line corresponds to the best fit $S(k)= 1/k^\alpha$ (the error in $\alpha$ is about $\Delta \alpha = \pm 0.02$) 
in the limit $N/k \gg 2 \pi b$ and the solid line is the prediction of WD statistics obtained from the GUE. The intersection 
 between the linear fit and the WD statistics prediction corresponds in principle with the Thouless energy of the 
system. However we observe that, 
for $b \leq 1$, $S(k)$ is always different from the WD prediction and consequently no clear Thouless energy can be defined.}  
\end{figure}

\section{ Analytical evaluation of $S(k)$. Power spectrum characterization of critical statistics.} 

Our goal is to compute the power spectrum 
\be 
\label{sk}
\langle S(k) \rangle =\left \langle \left| {1\over \sqrt{N}} \sum_{n=1}^N \delta_n \exp{\left(
{-2\pi i k n\over N} \right)} \right|^2 \right \rangle\;, 
\ee  
with  $\delta_n =\sum_{i=1}^n s_i-{\bar s}= \epsilon_{n+1}-\epsilon_1-n$, 
$\epsilon_n$ represents the nth unfolded eigenvalue of the critical random banded model Eq. (\ref{e4}) and 
the brackets stand for ensemble average.
 
In a first stage we evaluate $\langle S(k) \rangle $ in a continuous approximation, 
$ \langle S(k) \rangle = \langle S(k) \rangle_{cont} $, 
$\sum \rightarrow \int$, $s_i \rightarrow \rho(\epsilon)d\epsilon$ and 
${\bar s} \rightarrow  {\bar \rho(\epsilon')}d\epsilon'$ where 
$\rho(\epsilon)= \sum_i \delta(\epsilon - \epsilon_i)$ is the full 
 spectral density and
${\bar \rho(\epsilon)}$ is just the smooth part of it (the one utilized to unfold the spectrum).

The power spectrum is now given by,
\be
\langle S(k) \rangle_{cont} =\left \langle \left|\int d\epsilon \delta'(\epsilon) \exp{\left(
{-2\pi \epsilon k} \right)} \right|^2 \right \rangle = \left \langle \int_{-\infty}^{+\infty} \int_{-\infty}^{+\infty} d\epsilon'd\epsilon 
\exp(-2\pi i(\epsilon -\epsilon')k) \int^{\epsilon'} \int^{\epsilon} {\tilde \rho}(\alpha){\tilde \rho}(\alpha') d\alpha'
d\alpha \right \rangle 
\ee  
with $\delta'(\epsilon) = \int^{\epsilon}d\epsilon {\tilde \rho}(\epsilon)$ and 
${\tilde \rho}(\epsilon)=\rho(\epsilon)-{\bar \rho}(\epsilon)$.

After integrating by parts in $\epsilon$ and $\epsilon'$ the above expression simplifies to,
\be
\langle S(k) \rangle_{cont} =\frac{1}{4\pi^2 k^2}\int_{-\infty}^{+\infty} \int_{-\infty}^{+\infty}d\epsilon'd\epsilon 
\exp(-2\pi i(\epsilon -\epsilon')k)R_2(\epsilon,\epsilon')
\ee
where $R_2(\epsilon,\epsilon')$ is the TLCF defined previously. 
Since the spectrum is translational invariant (if we are far from the edges)
$R_2(\epsilon,\epsilon')=R_2(s=\epsilon-\epsilon')$ and
\be
\langle S(k) \rangle_{cont} =\frac{1}{4\pi^2k^2}\int_{-\infty}^{\infty}\exp(-2\pi i sk)R_2(s)= \frac{1}{4\pi^2k^2}K(k)
\ee
where $K(k)$, the Fourier transform of the TLCF, is usually referred to as the spectral form factor.

Once we have obtained an explicit expression for the power spectrum in 
 terms of known quantities as $R_2(s)$ we have to go back to the original discrete formulation. This 
 can be easily done by following standard relations between the discrete and the 
 continuous Fourier transform, we only present the final result and refer to \cite{four,rel1}
 for additional details,
\be
\label{sfi}
S(k) = \frac{K(t)}{4\pi^2 t^2} + \sum_{q=1}^{\infty}\frac{K(t+q)}{4\pi^2 (t+q)^2}+\frac{K(q-t)}{4\pi^2 (q-t)^2} + \beta 
\ee
with $t=k/N$ and $\beta$ a constant given by $\beta = [K(0) -1]^2/12$ which accounts for the differences between the 
 fluctuations of $\int^{\epsilon'} \int^{\epsilon} \langle {\tilde \rho}(\alpha){\tilde \rho}(\alpha') \rangle d\alpha'
d\alpha$ and those of the original discrete correlator $\delta_n$ (see \cite{four} for details). For the sake of simplicity
we set $\langle S(k) \rangle = S(k)$.  
The above expression combined with Eq \ref{r21},\ref{r23} 
provides with a closed and compact expression for the power spectrum as a function of the band size $b = 1/2\pi h$.

In the region of $b \gg 1$ the spectral form factor can be explicitly 
 evaluated by using the TLCF of Eq.(\ref{r21}). 
\be
K(t)=1 - \frac{1}{2}\left[ (1-t)\coth \left(\frac{2-2t}{h}\right)+  (1+t) \coth \left(\frac{2+2t}{h}\right)
-2t \coth \left (2t/h \right)\right]   
 \ee

We can distinguish two different regions. For $t \ll h =1/2\pi b$, corresponding 
 to eigenvalues separated a distance much larger than the mean level spacing, 
$K(t) \sim h/2$ is a constant and $S(k) \sim h/(8\pi^2 t^2)$,
similar to the case of Poisson statistics. However for Poisson $K(0)=1$ but in our case $K(0)= h/2$. 
This is an important difference
 since Poisson statistics is associated with eigenstates 
exponentially localized but for $K(0)= \chi \neq 1$ (where $\chi$ is the slope of the number variance)
 the eigenstates are multifractal. 

It seems that, at least in this case, the exponent of the decay of $S(k)$ does not completely 
 specify the nature of the quantum motion. 
We will go back to this point when we discuss applications of our work 
 in the context of quantum chaos.\\
 In the opposite limit $t \gg h = 1/2\pi b$, $K(t) = t$ and $S(t) \sim 1/(2\pi^2 t)$ 
 in agreement with the result for WD statistics (GUE).

The transition region separating the two types of decay 
($1/t$ and $1/t^2$) corresponds to the Thouless energy of the system. As usual it separates 
 short range correlations still controlled by WD statistics 
from larger scales in which typical features of the AT appear.

We have thus found that the power spectrum in the limit $b \gg 1$ corresponding to the case of a disordered 
 system in $2+\epsilon$ dimensions with short range disorder at the AT has different power-law decay depending of the 
spectral region of interest, these differences can be effectively utilized to find signatures of an AT from a given spectrum.

Analogously, in the region $b =1/2\pi h \ll 1$, which corresponds with the case of disordered conductor in 
 $d \gg 1$, a straightforward calculation 
shows that $K(t) = 1-1/\sqrt{2}h$ for $t \ll h$ and then goes to $K(t) = 1$ for $t \gg h$.
Consequently  $S(t) \sim 1/t^2$ up to scales smaller than the mean level spacing.  
Strictly speaking there is a narrow transition 
region already for $t > 1$ in which $K(t)$ is linear 
However it is difficult to interpret it as a Thouless energy 
since even for scales shorter than the mean level spacing the spectral correlations are different
 from WD statistics.

Finally we mention that for intermediate $b$ there is not a rigorous analytical relation 
 for the TLCF. However we shall see that the conjecture Eq (\ref{r23}) 
 describes very accurately the numerical results (see Fig 1). 
\vspace {1cm}
\begin{figure}[ht]
\hfill
  \begin{minipage}[t]{.45\textwidth}
      \includegraphics[width=\columnwidth]{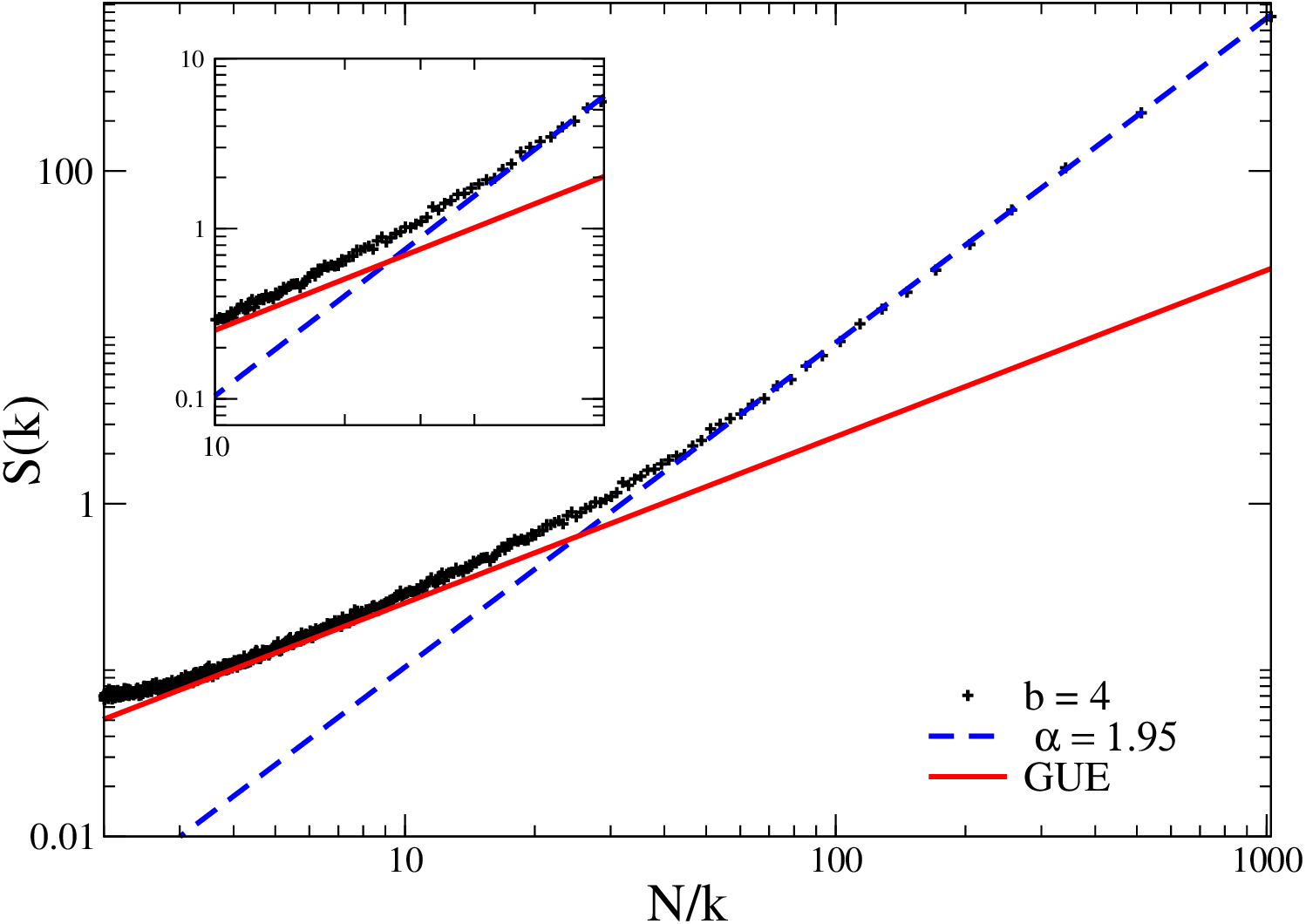}
      \label{f3}
  \end{minipage}
  \hfill
  \begin{minipage}[t]{.45\textwidth}
      \includegraphics[width=\columnwidth]{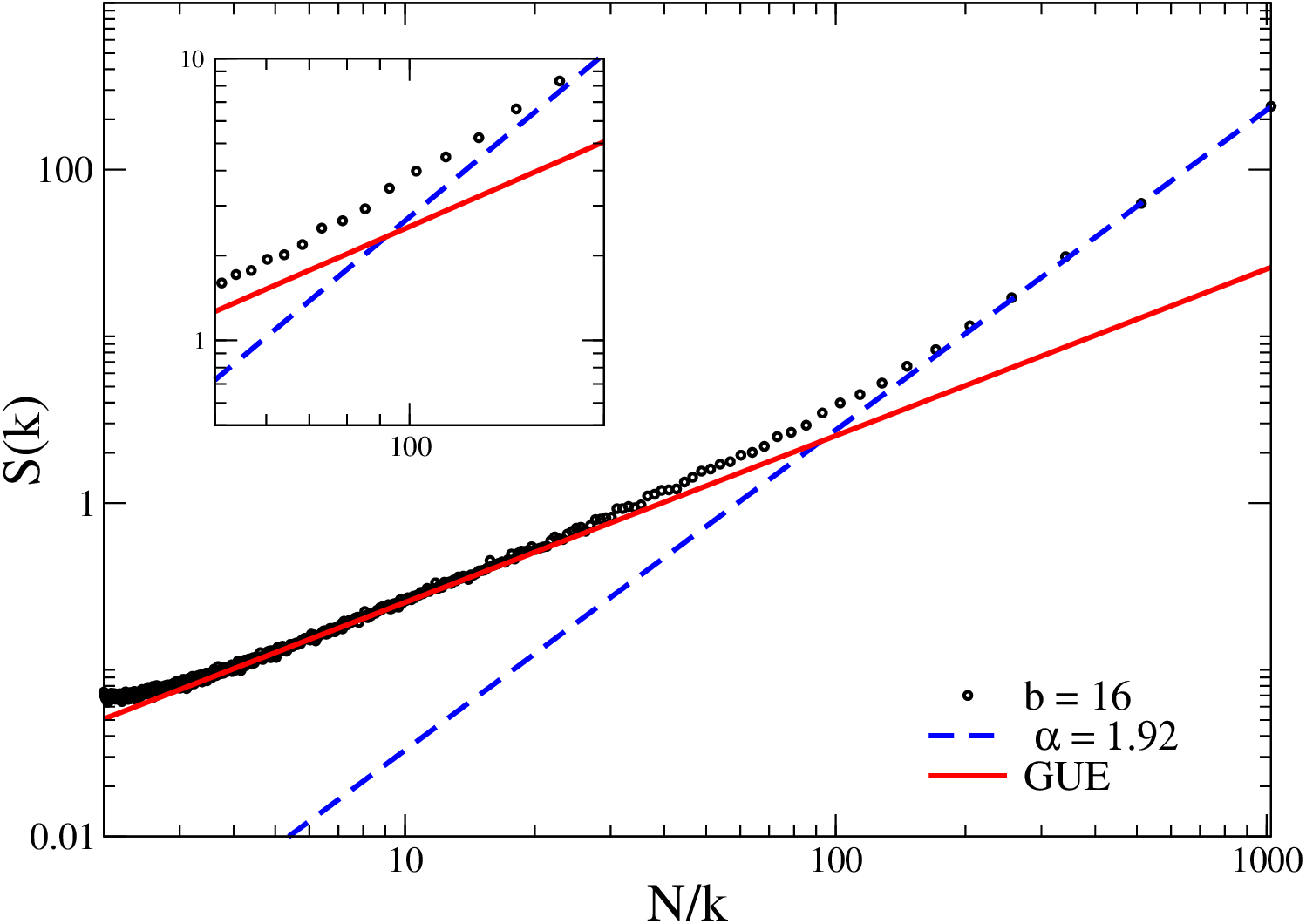}
  \end{minipage}
      \caption{Power spectrum $S(k)$ versus $N/k$ in the limit $b \gg 1$.
 Crosses and circles represent the numerical results (the matrix size is $3000$ and the 
 number of eigenvalues considered is $N=1024$ around the center of the band) 
for the critical random 
 banded model Eq.\ref{e4} for $b = 4,16$ respectively. The  power spectrum was evaluated
from Eq. \ref{sk}. The dashed line corresponds to the best fit $S(k)= 1/k^\alpha$ (the error in $\alpha$ is about $\Delta \alpha = \pm 0.02$) 
in the limit $N/k \gg 2 \pi b$ and the solid line is the prediction of WD statistics obtained from the GUE. The intersection 
 between the linear fit and the WD prediction corresponds with the Thouless energy of the 
system. The units has been chosen such that the intersection point  
 gives the dimensionless conductance of the system $g$. Thus 
$g \sim 25$ for $b = 4$ and $g \sim 100$ for $b = 16$.}
\end{figure}
\section{Numerical calculations}
We now investigate numerically the random banded model in Eq.\ref{e4} 
in order to test the analytical predictions of the previous section.

We studied the power spectrum $S(k)$ by   
  direct diagonalization of the critical random banded matrix Eq.\ref{e4}
for different matrix sizes (almost all of our plots are for $3000$ though 
 we try higher volumes, up to $5000$, in order to check that our results are not 
 size dependent).
 The 
 number of different realizations of disorder is chosen such that
 for each matrix size the total number of eigenvalues be at least $2 \times 10^6$. Typically 
around $35\%$ 
of the eigenvalues around the center of the band are utilized. 
The eigenvalues thus obtained 
 are unfolded with respect to the mean spectral density. 
The power spectrum is calculated by using Eq.\ref{sk} where the Fourier transform is evaluated by using a
fast Fourier transformation routine.

In Fig 1 we have plotted $\log S(k)$ for different $b$. In all cases the matrix size was $ 
3000$ and 
the evaluation of $S(k)$ was carried out within a band around the center of the spectrum 
 containing $1024$ eigenvalues. As observed the agreement
 between the analytical (with $R_2(s)$ given by Eq.(\ref{r23}) and $h =1/2\pi b$) and numerical results is excellent 
 for all $b$. Also in agreement with the analytical prediction we observe that, for $b \gg 1$,  
the power spectrum switches from $S(k) \sim 1/k$ for $N/K \ll h =1/2\pi b$ to 
 $S(k) \sim 1/k^2$ in the opposite limit. However for $b \ll 1$, $S(k) \sim 1/k^2$ 
for almost all accessible frequencies.\\
We conclude after the analytical and numerical analysis that the AT in a disordered conductor can be 
satisfactorily detected and examined by looking at the power spectrum of a signal $\delta_n$ consisting of the 
fluctuations around its mean value of the nearest neighboring spacings $s_i = \epsilon_{i+1}-\epsilon_i$.

We have also found that $S(k)$ provides with an accurate method to locate the Thouless energy of a generic
disordered conductor. As mentioned previously, the Thouless energy $E_c$ is a scale of energy related with the classical 
 diffusion time through the sample. In units of the mean level spacing $\Delta$ it gives the dimensionless conductance
 $g$, namely, the number 
 of eigenvalues $g = E_c/\Delta$ for which the universal results of WD statistics apply. 
From a practical point of view the evaluation of $g$ from 
 a given spectrum is a hard task since it may depend on what spectral correlator is used. Thus $\Sigma^2(L)$ gives a
prediction of $g$ bigger than that of $P(s)$ but much smaller 
than that of the spectral rigidity $\Delta_3 (L)$ (see \cite{mehta} for a definition).
Another problem is that even for each particular correlator the value of $g$ is somewhat ambiguous since it is far from clear how to locate even approximately the point in which WD ceases to be applicable.\\  
Below we show that $S(k)$ provides with a more efficient and precise way to locate and analyze $g$.  
The idea (see Fig 2 and Fig 3) is to plot $S(k)$ as a function of $N/k$. 
Then in the region $N/k \gg 2\pi b$ we fit $S(k)$ 
to a linear (in a log scale) curve $S(k) \sim 1/k^\alpha$ (from the previous analysis
 $\alpha \sim 2$). In the opposite limit $S(k)$ should be given by the prediction of WD statistics. 
We define the Thouless energy as the intersection between the 
WD prediction and the linear fit.

In Fig 2 we see that for $b \gg 1$ the intersection of these two curves gives $g \sim  2\pi b \gg 1$  
   in good agreement with the theoretical prediction. However in the region $b \leq 1$ (see Fig 3), though 
 formally a Thouless energy can be defined thorough the intersection of the two curves, 
its interpretation as the  limit of applicability of WD statistics is dubious since even in the limit $N/k \ll 2\pi b$ 
deviations with respect to the WD prediction are clearly visible.

Finally we mention that, as observed in Fig 1, the best fit of the numerical
results does not occur at the analytical estimation $h = 1/2\pi b$. 
There are two reasons for that disagreement:
The analytical results are strictly valid only at the center of the band. Eigenstates beyond this region 
are still critical but are described by an effective bandwidth \cite{cambridge} smaller than $b$. On the other hand finite 
 size effects are important
 in the limit $t \rightarrow 0$ since we are testing the largest eigenvalue separations. 
However we have decided not to reduce the spectral window in order to give a full global picture of the 
power spectrum at the AT. After all, as shown in Fig 1, 
these effects are easily compensated by slightly modifying $b$.
\vspace{1cm}
\begin{figure}[ht]
  \hfill
  \begin{minipage}[t]{.45\textwidth}
      \includegraphics[width=\columnwidth]{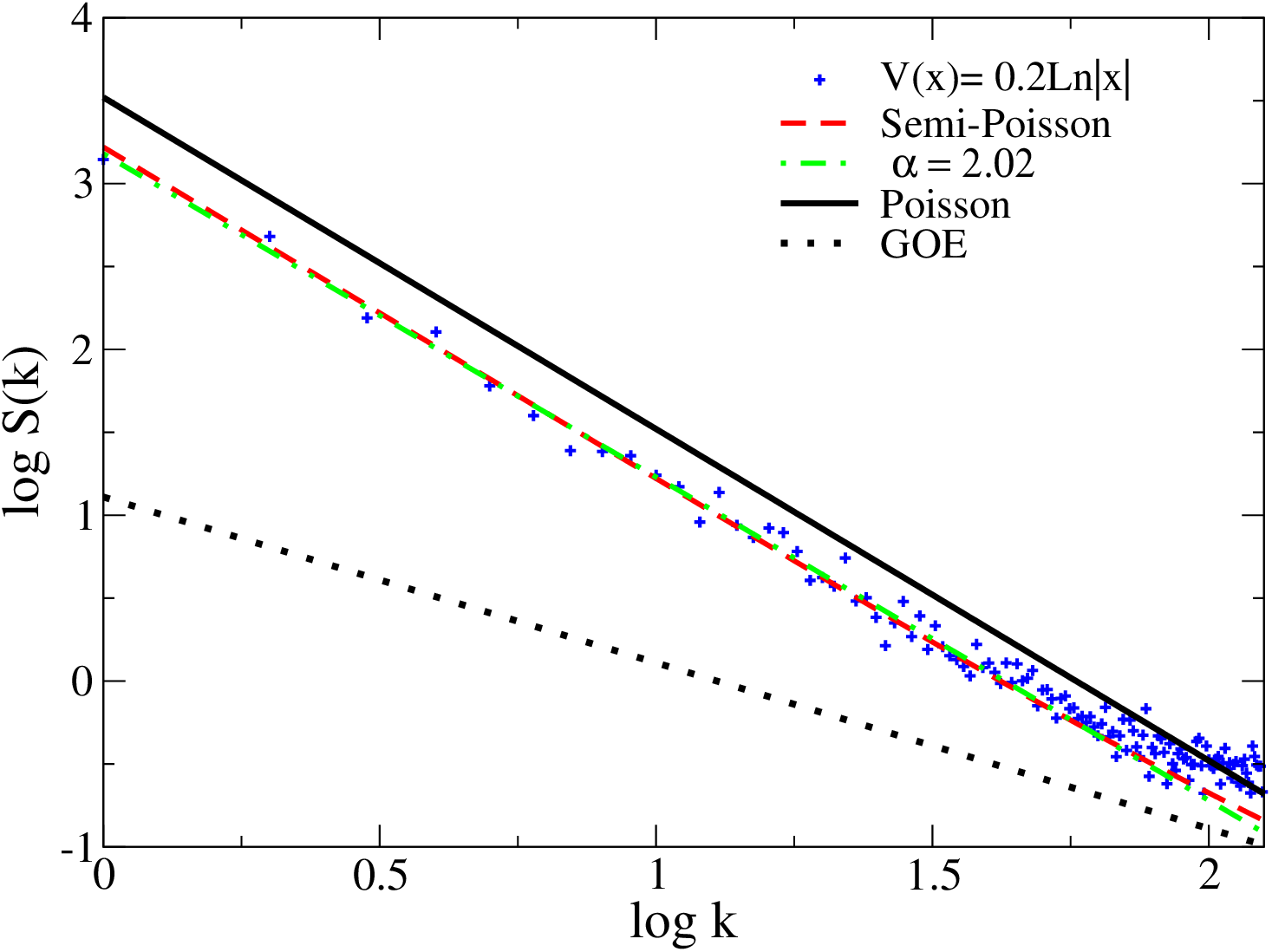}
      \label{f4}
  \end{minipage}
  \hfill
  \begin{minipage}[t]{.45\textwidth}
      \includegraphics[width=\columnwidth]{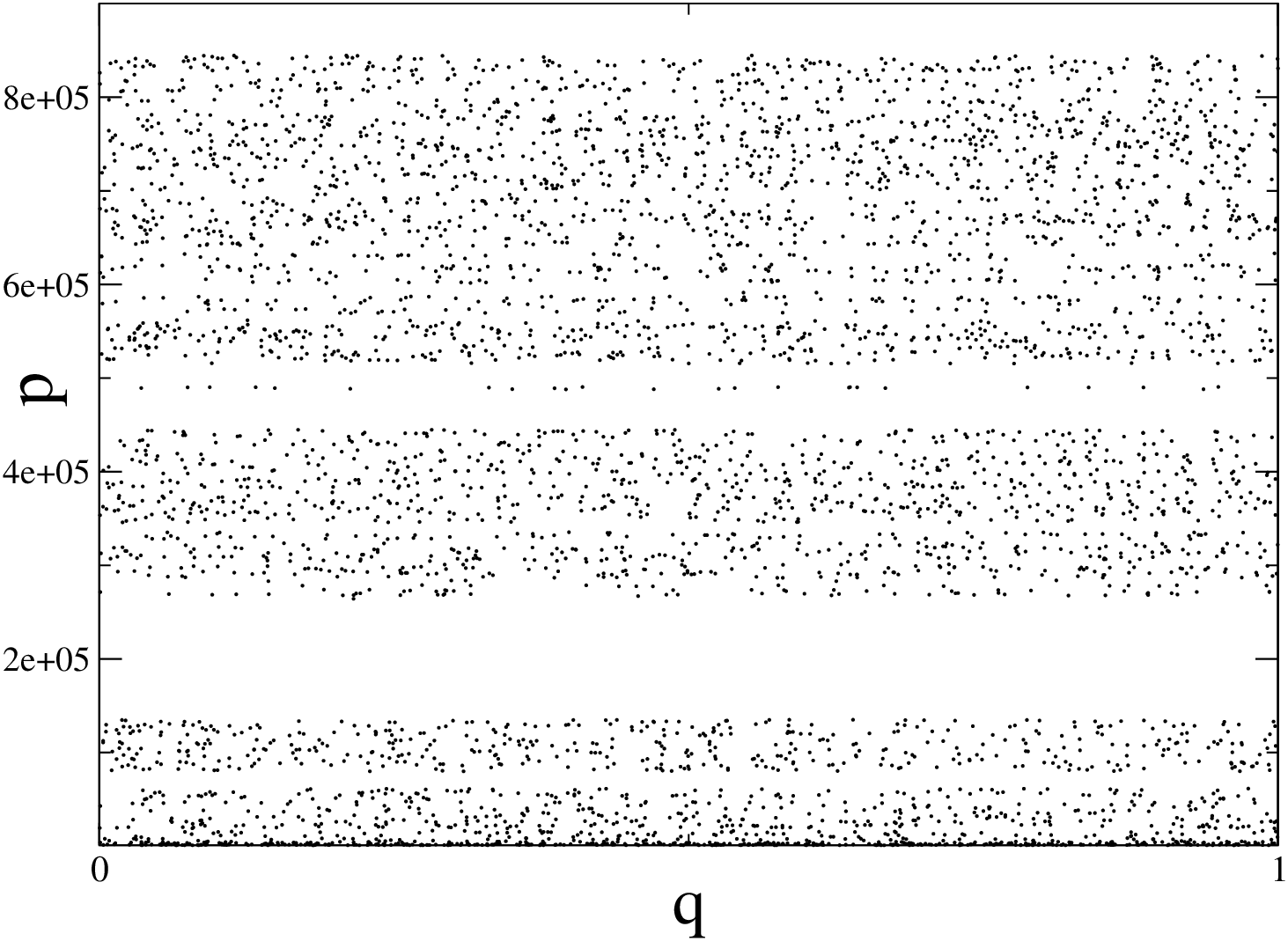}
  \end{minipage}
  \hfill
      \caption{(Left) Power spectrum $S(k)$ obtained from $25$ set of $256$ eigenvalues
 after the numerical diagonalization
 of the evolution 
 matrix associated to  Eq. (\ref{ourmodel}) with  
$V(q)=0.2Ln |q|$ (cross). The dashed line is the prediction of SP statistics, the dotted-dashed line
 is the best fit $S(k) \sim 1/k^{\alpha}$ and the dotted line is the prediction of WD statistics 
 as obtained from the Gaussian Orthogonal  Ensemble (GOE) of random matrices. (Right) 
   Poincare section from a single initial condition $p_0 =0.2$ and $q_0 = 0.6$ after 300000 iterations.  
} 
\end{figure}
\section{Application to quantum chaos}
In this final section we investigate possible applications of our previous findings 
in the context of quantum chaos and also argue that $S(k) \sim 1/k^2$  is not exclusive of
 quantum deterministic system whose classical counterpart is integrable. 

Critical statistics and multifractal wavefunctions are very much universal so 
they should also appear in deterministic quantum systems.
Indeed, in a recent letter \cite{ant9} we have established a novel relation 
 between the presence of anomalous diffusion in the classical dynamics, the 
singularities of a classically chaotic potential and the power-law localization of the 
 quantum eigenstates. Specifically, it was found that for a kicked rotor with Hamiltonian 
\be
\label{ourmodel}
{\cal H}= \frac{p^2}2 +V(q)\sum_n\delta(t -nT)
\ee
(with $q \in [-1,1)$) 
 both level statistics and eigenfunctions are similar to the ones at the AT (critical statistics)  
provided that $V(q)$ has a $\log$ (in the simplest case $V(q)=\epsilon_0 Ln |x|$)
or a step-like \cite{ant11} singularity.
It was 
 also found in \cite{ant9} that these findings are universal in the 
 sense that  neither the classical
nor the quantum properties depend on the details of the potential but only on the
type of singularity.
Deviations from WD statistics not coming from a mixed phase space has also been 
 reported in a variety of systems:  
Coulomb billiard\cite{altshu}, Anisotropic Kepler problem \cite{wintgen},  
a kicked rotor in a well potential \cite{bao} and pseudointegrable billiards \cite{bogo3,bogo04}.
For the latter it was found  {\cite{bogo3}} that the level statistics is accurately 
 described by a 
the classical Dyson gas with the logarithmic 
pairwise interaction restricted to a finite number $k$ of nearest neighbors. 
 Analytical solutions are available for general $k$. 
 For $k=2$, usually referred to as semi Poisson (SP) statistics,
 $R_2(s)=1-e^{-4s}$, $P(s)=4s e^{-2s}$ and $\Sigma^2(L) = L/2 +(1-e^{-4L})/8$. 
 We have also found \cite{ant11} that SP statistics describes accurately the spectral
 correlations of the above kicked rotor with a 
 step like singularity and also provides with a reasonable description for the 
$\log$ singularity but only for $\epsilon_0 \sim 0.2$. 

Due to the simplicity of the TLCF in SP statistics one can evaluate the power spectrum exactly,
\be
S(k)=\frac{1}{4\pi^2 t^2}\left[1- \frac{8}{16+4\pi^2t^2} \right]
\ee

Thus the power spectrum associated to SP statistics has $1/t^2$ decay even though the classical dynamics 
 is not integrable, this is also in agreement with the prediction of critical statistics for $b \ll 1$. 
Thus $S(k) \sim 1/t^2$ ($t=k/N$) is not always a signature of classical integrable dynamics.
 Although generically the power spectrum associated 
with classically integrable systems  has this feature, other types of non integrable 
 dynamics may have $1/t^2$ as well. In order to fully characterize the classical dynamics from $S(k)$ one has to   
 specify not only the exponent but also and additional point of the curve, for instance $K(0)$.  
The point is that a $1/t^2$ decay only tell us that the form factor is constant. However, as mentioned previously, the  
physical properties of the system are strongly modified by a spectral form factor different from the unity.

As a further corroboration of our claims we have evaluated numerically $S(k)$ for the Hamiltonian 
of Eq. \ref{ourmodel} with a potential $V(x)=0.2 Ln |x|$. We diagonalize numerically 
the evolution matrix associated to the Hamiltonian Eq.\ref{ourmodel} for $N=5200$, 
$S(k)$ is obtained from Eq.\ref{sk}. In order to improve statistics we divide the 
 original spectrum in 20 set of 256 eigenvalues.
 As shown in Fig 4, $S(k) \sim 1/t^2$ for almost all $t$ in close agreement 
 with the prediction of SP or critical statistics. 
However (see right plot) the associated Poincare section obtained from
 just a single initial condition 
is very different from that of an system whose classical dynamics is integrable.

\section{Conclusions}

We have shown that the power spectrum of the energy level fluctuations  
 at the Anderson transition for a family of critical power-law random banded 
 matrices is characterized by a power spectrum which 
 $1/f^2$ noise for small frequencies and $1/f$ noise for larger frequencies.
  In the weak disorder limit $b \gg 1$, the analysis 
 of the transition region between these two power-law limits provides with an accurate estimation of the  
 Thouless energy of the system. 
As disorder increases the Thouless energy looses its meaning and the 
 power spectrum presents a $1/f^2$ noise up to frequencies related to the Heisenberg time of the system.
Finally we discuss under what circumstances these findings 
 may be relevant in the context of non-random Hamiltonians. Specifically it is shown
that the exponent of the power-law decay of $S(k)$
 does not fully specify the type of motion of the classical counterpart.

\vspace{1cm}
We acknowledge financial support from a postdoctoral fellowship of Spanish Ministry of Science 
and Education.


\end{document}